\documentclass[8pt]{iopart}
\usepackage{iopams} 
\usepackage[latin1]{inputenc}
\usepackage{graphicx}
\usepackage{url}

\begin{document} 
 
\title[Spontaneous emission -- graphene waveguides -- surface plasmon polaritons]{Graphene coated  subwavelength  wires: A theorical investigation of the emission and radiation properties}
\author{Mauro Cuevas} 
\address{Consejo Nacional de
Investigaciones Cient\'ificas y T\'ecnicas (CONICET) and Facultad de Ingenier\'ia y Tecnolog\'ia Inform\'atica, Universidad de Belgrano,
Villanueva 1324, C1426BMJ, Buenos Aires, Argentina}
\address{Grupo de Electromagnetismo Aplicado, Departamento de F\'isica, FCEN, Universidad de Buenos Aires,  Ciudad Universitaria,
Pabell\'on I, C1428EHA, Buenos Aires, Argentina}
\ead{cuevas@df.uba.ar}

\begin{abstract} 
This work analyzes the  emission and radiation properties of a single optical emitter embedded  in a graphene--coated subwavelength wire.  We discuss the modifications of the spontaneous emission rate and the radiation  efficiency as a function of the position  and orientation of the dipole inside the wire. Our results show that these quantities can be  enhanced by several orders of magnitude when the emission frequency coincides with one of the  resonance frequencies of the graphene--coated wire. In particular,  high--order     plasmon resonances are excited when the emitter is moved from the wire center. The modifications by varying the orientation of the dipole in the  near field distribution and in the far field intensities are shown.
\end{abstract} 



\maketitle
\ioptwocol

\section{Introduction} 

Surface plasmon polariton results from the coherent coupling of photons to  surface charge density  oscillations \cite{raether,maier}. 
In natural plasmonic materials such as metals,  
constitutive parameters such as the conductivity and charge density are fixed, whereas they can be tuned in graphene electrically or by chemical doping \cite{Xia},  which has a dramatic effect on its optical
properties. As the linear band structure of graphene causes the plasmon mass to depend on the Fermi--level position, electrically tunable surface plasmons can be supported by graphene from microwaves to the mid--infrared regimes \cite{jablan}. This has aroused new interest in studying graphene in the context of optical and plasmonic applications,  
including graphene quantum dots as a new generation of light--emitting devices \cite{GQD}, sensing \cite{francisco}, solar cells \cite{bonaccorso}, to mention just a few. 

Surface plasmons polaritons can be roughly divided into two categories: surface plasmon (SPs) propagating along waveguiding structures, such as an infinite flat graphene monolayer, and localized surface plasmons (LSPs) supported by spatially limited structures, such as
scattering particles.  
Both kind of plasmon modes can be excited when a single optical emitter (such as quantum dots and single molecules) is placed next to one of the aforementioned structures.  As a result of the high light confinement, an  enhanced decay rate of the emitter into the plasmonic mode via the Purcell effect  take place \cite{maier_nature}.
In fact, it was recently showed that  the interplay between an optical emitter and a graphene--coated  sphere leads to an enhancement of the spontaneous emission into confined plasmonic modes \cite{LSP}.  
A variety of structures such as infinite graphene monolayers \cite{grafeno,grafeno1,grafeno2,grafeno4}, ribbons or nanometer sized disks \cite{grafeno3} and double graphene waveguides \cite{guia1,cuevas1} have been the object of intensive research over the last few years due to the possibility to engineer surface plasmon  mode density of states   
to control  emission properties.

In this paper we consider a cylindrical dielectric core coated with a graphene layer   
and we investigate the role of the LSPs in modifying the emission and radiation rates of an emitter placed inside the graphene--coated cylinder. In this context, several works focused on the influence that  the eigenmodes play on circular  dielectric and metallic waveguides \cite{mcphedran,dereux,marocico} or on single wall carbon nanotubes \cite{martin_moreno}. Such systems 
offer a new tool for the interplay between light and matter at the nanometer scale \cite{soresen,akimov}. 
 Other works have presented two dimensional calculations that extend such studies to examining the role of the shape of the waveguide \cite{rogobete1,rogobete2}. 
Two dimensional setting has important properties that allow to understand better the three dimensional case \cite{mcphedran,sondergaard}, since qualitatively similar trends will hold in three dimensions \cite{rogobete2}. 
Experimental advances in the fabrication of micro--structures based on insulating materials such as SiO$_2$ doped with  
molecular ions \cite{vidrio},  
the possibility of encapsulate single atom, molecules and compounds into graphene wires \cite{encapsulado1}  and the fact that thanks to the van der Waals force, a graphene sheet can be tightly coated on a fiber surface \cite{coated}, 
 encourages an  investigation about  the emission and radiation characteristics of such graphene based systems.     

The paper is organized as follows. First, in Section \ref{teoria} we sketch an analytical method based on the separation of variables approach and obtain a solution for the electromagnetic field scattered 
 by a graphene--coated wire when an oscillating  line dipole source is located at an arbitrary position inside the  wire cylinder. We derive analytical expressions for the power emitted and radiated by the source.  In section \ref{resultados} we present examples of emission and radiation decay rates  corresponding to wires tightly coated with a graphene layer and  compared the results with those obtained in the absence of the graphene coating. Finally,  concluding remarks are provided in Section \ref{conclusiones}. 
The Gaussian system of units is used
 and an $\mbox{exp}(-i\, \omega\, t)$ time--dependence is implicit throughout the paper, with $\omega$ as the angular frequency, $t$ as the time, and $i=\sqrt{-1}$. The symbols Re and Im are used for denoting the real and imaginary parts of a complex quantity, respectively.

\section{Theory} \label{teoria} 

We consider a graphene coated cylinder with circular cross section (radius $a$) centered at $x=0$, $y=0$ (figure 1). The wire substrate is characterized by the electric permittivity $\varepsilon_1$ and the magnetic  permeability $\mu_1$. The coated wire is embedded in a transparent medium with electric permittivity $\varepsilon_2$ and magnetic permeability $\mu_2$. 
The graphene layer is considered as an infinitesimally thin, local and isotropic two--sided layer with frequency--dependent surface conductivity $\sigma(\omega)$ given by the Kubo formula \cite{falko,milkhailov}, which can be read as  $\sigma= \sigma^{intra}+\sigma^{inter}$, with the intraband and interband contributions being
\begin{equation} \label{intra}
\sigma^{intra}(\omega)= \frac{2i e^2 k_B T}{\pi \hbar (\omega+i\gamma_c)} \mbox{ln}\left[2 \mbox{cosh}(\mu_c/2 k_B T)\right],
\end{equation}  
\begin{eqnarray} \label{inter}
\sigma^{inter}(\omega)= \frac{e^2}{\hbar} \Bigg\{   \frac{1}{2}+\frac{1}{\pi}\mbox{arctan}\left[(\omega-2\mu_c)/2k_BT\right]-\nonumber \\
   \frac{i}{2\pi}\mbox{ln}\left[\frac{(\omega+2\mu_c)^2}{(\omega-2\mu_c)^2+(2k_BT)^2}\right] \Bigg\},
\end{eqnarray}  
where $\mu_c$ is the chemical potential (controlled with the help of a gate voltage), $\gamma_c$ the carriers scattering rate, $e$ the electron charge, $k_B$ the Boltzmann constant and $\hbar$ the reduced Planck constant.
The intraband contribution dominates for large doping $\mu_c<<k_BT$ and is a generalization of the Drude model for the case of arbitrary band structure, whereas the interband contribution dominates for large frequencies $\hbar \omega \geq \mu_c$. 
A line dipole source (whose axis lie along the $\hat{z}$ axis) with a  dipole  moment $\vec{p}=p (\cos\alpha\, \hat{x}+\sin\alpha \, \hat{y})$ is placed inside the cylinder, at position $\vec{r^{'}}=\rho^{'} \hat{r}+\phi^{'}  \hat{\phi}$ ($\rho^{'}<a$).  The dipole is aligned at an angle $\alpha$ with respect to the $\hat{x}$ axis, as indicated in figure 1.  
 The current density of the electric dipole  is
\begin{eqnarray}\label{corriente}
\vec{j}(\vec{r})=-i \omega \vec{p} \, \delta(\vec{r}-\vec{r}')=-i \omega \vec{p} \, \frac{1}{\rho} \delta(\rho-\rho^{'}) \delta(\phi-\phi^{'}).
\end{eqnarray} 
In an unbounded medium, the dipole fields are obtained from the vector potential $\vec{A}$ (refer to appendix for its derivation),
\begin{eqnarray}\label{A}
\vec{A}(\rho,\phi)=\sum_{m=-\infty}^{+\infty} \pi k_0 J_m(k_1\rho_<)H_m^{(1)}(k_1\rho_>)\,e^{i m (\phi-\phi^{'})} \\ \nonumber
\times \left[p_{\rho} \hat{r}+ p_\phi \hat{\phi}\right],
\end{eqnarray} 
\begin{eqnarray}\label{H}
\vec{H}(\rho,\phi)=\nabla \times \vec{A}=\hat{z} \varphi(\rho,\phi),
\end{eqnarray} 
\begin{eqnarray}\label{E}
\vec{E}(\rho,\phi)=\frac{i}{k_0 \varepsilon_1}\nabla \times \vec{H}(\rho,\phi)=-\frac{i}{k_0 \varepsilon_1} \hat{z} \times \nabla_t \varphi,
\end{eqnarray}
where $\varphi(\rho,\phi)$ is the non--zero component of the total magnetic field along the axis of the wire ($\hat{z}$ axis), $\nabla_t=\hat{\mbox{r}}\frac{\partial}{\partial \rho}+\hat{\phi} \frac{1}{\rho} \frac{\partial}{\partial \phi}$ is the transverse part of the $\nabla$ operator, 
$k_0=\omega/c$ is the modulus of the photon wave vector in vacuum, $\omega$ is the angular frequency, $c$ is the vacuum speed of light, $\rho_<$ ($\rho_>$) is the smaller (larger) of $\rho$ and $\rho^{'}$,  $p_\rho$ and $p_\phi$ are the projection of vector $\vec{p}$ on the $\hat{r}$ and $\hat{\phi}$ axis, respectively, and $J_m$ and $H_m^{(1)}$ are the nth Bessel and Hankel functions of the first kind, respectively. 
From Eqs. (\ref{A}) and (\ref{H}), we obtain the primary magnetic field emitted by the dipole,
\begin{eqnarray}\label{inc1}
\varphi_{i}(\rho,\phi) = \sum_{m=-\infty}^{+\infty} \pi k_0 k_1 J_m(k_1\rho^{'})\\ \nonumber
\times \left[H_m^{(1)'}(k_1\rho)\,p_\theta-i m \frac{H_m^{(1)}(k_1\rho)}{k_1\rho}\,p_\rho\right]
 e^{i m (\phi-\phi^{'})},
\end{eqnarray} 
for $\rho>\rho^{'}$, and
\begin{eqnarray}\label{inc2}
\varphi_{i}(\rho,\phi)=\sum_{m=-\infty}^{+\infty} \pi k_0 k_1 H_m^{(1)}(k_1\rho^{'}) \\ \nonumber
\times \left[J_m^{'}(k_1\rho)\,p_\theta-i m \frac{J_m(k_1\rho)}{k_1 \rho}\,p_\rho\right] 
 e^{i m (\phi-\phi^{'})},
\end{eqnarray} 
for $\rho<\rho^{'}$.
 
When the dipole is located inside the coated wire cylinder, the scattered magnetic field along the axis of the wire, \textit{i.e}, the $\hat{z}$ component, denoted by $\varphi^{(j)}_s$ ($j=1,\,2$), are expanded as a series of cylindrical harmonics, one for the internal region ($\rho < a$, superscript $1$) and another one for the external region ($\rho > a$, superscript $2$),
\begin{eqnarray}\label{phi1}
\varphi^{(1)}_s(\rho,\phi)=\sum_{m=-\infty}^{+\infty} a_m J_m(k_1 \rho) e^{i m \phi}, 
\end{eqnarray}
\begin{eqnarray}\label{phi2}
\varphi^{(2)}_s(\rho,\phi)=\sum_{m=-\infty}^{+\infty} b_m H_m^{(1)}(k_2 \rho) e^{i m \phi}, 
\end{eqnarray}
where $a_m$ and $b_m$ are unknown complex coefficients.  
Taking into account that because of the graphene coating the tangential components of the magnetic field  are no longer continuous across the boundary --as they were in the case of uncoated cylinders-- the boundary conditions for our case can be expressed as
\begin{eqnarray}\label{cc1}
\frac{1}{\varepsilon_1}\frac{\partial}{\partial \rho}(\varphi_{i}+\varphi^{(1)}_s)|_{\rho=a}=\frac{1}{\varepsilon_2}\frac{\partial}{\partial \rho}\varphi^{(2)}_s|_{\rho=a},
\end{eqnarray} 
and 
\begin{eqnarray}\label{cc2}
\varphi^{(2)}_s|_{\rho=a}-(\varphi_{i}+\varphi^{(1)}_s)|_{\rho=a}=\\ \nonumber
\frac{4 \pi \sigma}{c k_0 \varepsilon_1}i \frac{\partial }{\partial   \rho} (\varphi_{i}+\varphi^{(1)}_s)|_{\rho=a}.
\end{eqnarray} 
Inserting the expressions (\ref{phi1}), (\ref{phi2}) and (\ref{inc1}) into the boundary conditions, we obtain the amplitudes of the scattered fields,
\begin{eqnarray}\label{am}
a_m=\frac{\pi k_0 }{2 i a J_m(x_1)\,D_m}\\ \nonumber
\times [k_1 a A_m h_m(x_2)+\frac{4\pi\sigma}{c}i k_0 a h_m(x_2) B_m-B_m)],
\end{eqnarray}
\begin{eqnarray}\label{bm}
b_m=\frac{\pi k_0}{2 i a H^{(1)}_m(x_2)\,D_m}[k_1 a A_m j_m(x_1)-B_m],
\end{eqnarray}
where
\begin{eqnarray}\label{dm}
D_m=h_m(x_2)-j_m(x_1)+\frac{4\pi\sigma}{c}i k_0 a j_m(x_1) h_m(x_2),
\end{eqnarray}
\begin{eqnarray}\label{jm}
j_m(x_1)=\frac{J_m^{'}(x_1)}{x_1 J_m(x_1)}, 
\end{eqnarray}
\begin{eqnarray}\label{jm}
h_m(x_1)=\frac{H_m^{(1)'}(x_2)}{x_2  H_m^{(1)}(x_2)}, 
\end{eqnarray}
\begin{eqnarray}\label{Bm}
A_m=p [J_{m-1}(k_1 \rho^{'})f_{m-1}(x_1) 
e^{i [\phi'-\alpha]}- \\ \nonumber
 J_{m+1}(k_1 \rho^{'})  g_{m+1}(x_1) 
e^{-i [\phi'-\alpha]}] 
e^{-i m \phi'}
\end{eqnarray}
\begin{eqnarray}\label{Am}
B_m=\frac{\partial}{\partial x_1}A_m =       \\ \nonumber
p [J_{m-1}(k_1 \rho^{'}) f_{m-1}'(x_1) 
 e^{i[\phi'-\alpha]} - \nonumber\\
 J_{m+1}(k_1 \rho^{'}) g_{m+1}'(x_1)e^{-i [\phi'-\alpha]}]  e^{-i m \phi'}, 
\end{eqnarray}
\begin{eqnarray}
f_m(x_1)=H_{m}^{(1)'}(x_1)-m\,\frac{H_{m}^{(1)}(x_1)}{x_1},\\  \nonumber
g_m(x_1)=H_{m}^{(1)'}(x_1)+m\,\frac{H_{m}^{(1)}(x_1)}{x_1}
\end{eqnarray}
and $x_1=k_1 a$, $x_2=k_2 a$. Here, the prime denotes the derivative with respect to the argument. 
The denominator $D_m$ in Eqs. (\ref{am}) and (\ref{bm}) are the same as that of the Mie scattering coefficients and may become very small in magnitude when the emission frequency coincides with one of the resonance frequencies \cite{RCD,CRD}.  

\begin{figure}[htbp]
\centering
\resizebox{0.40\textwidth}{!}
{\includegraphics{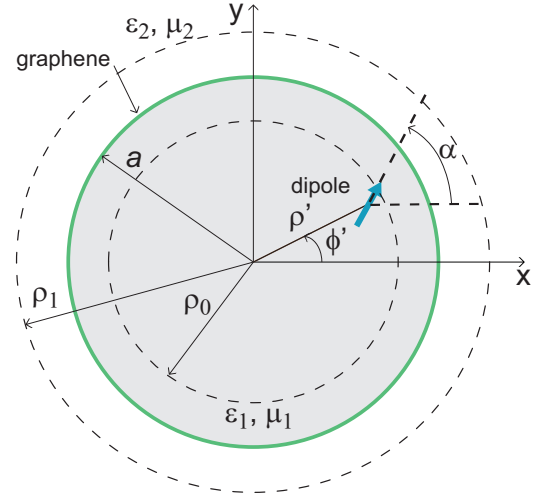}}
\caption{\label{fig:epsart} Schematic illustration of the system. An optical dipole emitter is inside a graphene--coated dielectric cylinder. The cylinder radius is $a$ and the graphene  surface conductivity is $\sigma$.}\label{sistema}
\end{figure}

In the limit of $\rho^{'}\rightarrow 0$, $A_m=H_{m}^{(1)'} p (e^{-i \alpha} \delta_{m,1}-e^{i \alpha} \delta_{m,-1})$, $B_m=H_{m}^{(1)''} p (e^{-i \alpha} \delta_{m,1}-e^{i \alpha} \delta_{m,-1})$ and, as a consequence the amplitudes $a_m$ and $b_m$ are zeros except $a_{\pm 1}$ and $b_{\pm 1}$, as expected from symmetry arguments.   
%
%
Once the amplitudes are determined, the scattered field, given by Eqs. (\ref{phi1}) and (\ref{phi2}) can be calculated at every point in the interior region (medium 1) and in the exterior region (medium 2). The total electromagnetic field allows us to calculate optical characteristics such as the power emitted and the power  radiated by the dipole.  The time average emitted power can be calculated from the integral of the normal component of the complex Poynting vector flux through an imaginary cylinder of length $L$ and radius $\rho_0<a$ that encloses the dipole (see Figure \ref{sistema})
\begin{eqnarray}\label{P}
P= \rho_0\,L\int_0^{2 \pi} \mbox{Re} \left\{\vec{S}(\rho_0,\phi)\cdot\hat{\mbox{r}} \right\} d\phi,
\end{eqnarray}
where
\begin{eqnarray}\label{S}
\vec{S}(\rho_0,\phi)= \frac{c}{8\pi} \left(-\frac{i}{k_0 \varepsilon_1}\right)\hat{z}\times \nabla_t [\varphi_i+\phi^{(1)}_s] \\ \nonumber
 \times \hat{z}[\varphi_i+\phi^{(1)}_s]^{\ast}.
\end{eqnarray}
Introducing Eq. (\ref{S}) into Eq. (\ref{P}), we obtain
\begin{eqnarray}\label{P1}
P= \\ \nonumber
 \frac{ \rho_0\,L\,c^2}{8 \pi \omega \varepsilon_1} \int_0^{2 \pi} \mbox{Re} \left\{ -i\, [\varphi_i+\varphi^{(1)}_s]^{*} \frac{\partial}{\partial \rho}[\varphi_i+\varphi^{(1)}_s] \right\} d\phi.
\end{eqnarray}
After some algebraic manipulation, the following results are found (see appendix) 
\begin{eqnarray}\label{P2}
P= \frac{\pi \omega^3\,p^2}{4\,c^2} +
\frac{ \omega \,p}{4 \sqrt{\varepsilon_1}}      \\ \nonumber
\times  \mbox{Re}\left\{i \sum_{m=-\infty}^{+\infty} J_m(k_1\rho^{'})\left[a_{m+1}\, e^{i  \alpha}+a_{m-1}\, e^{-i  \alpha}\right] e^{i m \phi^{'}} \right\}. 
\end{eqnarray}
Similarly, the time average radiative power emitted can be  evaluated by calculating the complex Poynting vector flux through an imaginary cylinder of length $L$ and radius $\rho_1>a$ that encloses the graphene--coated cylinder  (see Figure \ref{sistema})
\begin{eqnarray}\label{Pr}
P_{sc}=\frac{ \rho_1\,L\,c^2}{8 \pi \omega \varepsilon_2} \int_0^{2 \pi} \mbox{Re} \left\{ -i\, \left[\varphi^{(2)}_s\right]^{*} \frac{\partial}{\partial \rho} [\varphi^{(2)}_s] \right\} d\phi,
\end{eqnarray}
In the far--field region the calculation of the scattered fields given by  Eq. (\ref{phi2}) can be greatily simplified using the asymptotic expansion of the Hankel function for large argument. After some algebraic manipulation, we obtain
\begin{eqnarray}\label{Pr2}
P_{sc}=\frac{ \rho_1\,L\,c^2}{2 \pi \omega \varepsilon_2} \sum_{m=-\infty}^{+\infty} |b_m|^2.
\end{eqnarray}
To characterize quantitatively the effect of the  graphene coating, we define the normalized spontaneous  emission rate  $F$ as the ratio between the power emitted by the dipole, given by Eq. (\ref{P2}), and the power emitted by the same dipole  embedded in an unbounded medium 1. 
In a similar way,
 the radiative efficiency $F_{sc}$ is defined as the ratio between the power radiated by the dipole, given by Eq. (\ref{Pr2}), and the power  emitted by the dipole in the unbounded medium 1. 

\section{Results}\label{resultados}

We first briefly illustrate the emission and radiation characteristics in a dielectric wire without graphene coating (bare wire). Next, we consider a dielectric  wire wrapped with a graphene coating. In all the examples the core (radius $a=0.5\mu$m) is made of a transparent material ($\varepsilon_1 = 3.9$, $\mu_1=1$)  and is embedded in vacuum ($\mu_2 = \varepsilon_2 = 1$).    Rotational symmetry allows the emitter to be positioned on the $\hat{x}$ axis ($\phi'=0$).

\subsection{Uncoated dielectric wire}

\begin{figure}
\centering
\resizebox{0.450\textwidth}{!}
{\includegraphics{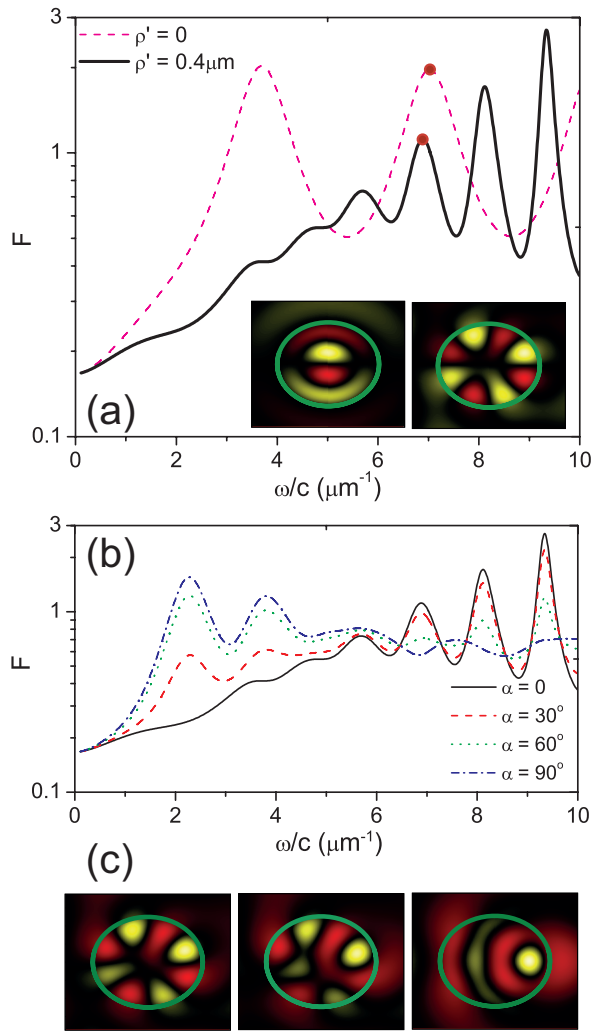}}
\caption{\label{fig:epsart} Spontaneous emission efficiency for a bare dielectric wire with a radius  $a=0.5\mu$m, constitutive parameters  $\varepsilon_1=3.9$ and $\mu_1=1$ in a vacuum, calculated for $\alpha=0^\circ$ (a), and for $\alpha=0^\circ,\,30^\circ,\,\,60^\circ,\,\,90^\circ$ (b). The insets in (a) show the magnetic field distributions $\varphi(\rho,\phi)$ near the wire for $\omega/c=7\mu$m$^{-1}$. Red, negative values, yellow, positive values. (c) Near magnetic  field distribution for the octupole resonance, $\alpha=30^\circ,\,60^\circ,\,90^\circ$ and $\rho'=0.4\mu$m. 
}\label{figura_sg}
\end{figure}

 In Figure \ref{figura_sg}a we plot the frequency dependence of the normalized  spontaneous emission  (per unit length) for an  emitter localized at  $\rho'=0$, on the wire axis, and for $\rho'=0.4\mu$m. The dipole moment $\vec{p}$ is oriented in the radial direction ($\alpha=0$). 
Due to the dielectric wire is lossless,  
the curve  in this figure  also represents the radiation decay rate efficiency of the emitter. 
The  peaks on the curve for $\rho'=0$ correspond to the first two Mie dipolar resonances and are associated with the complex  poles at the zeros of the denominators $D_1$  
of the coefficients $a_1$ and $b_1$ in Eqs. (\ref{am}) and (\ref{bm}) with $\sigma=0$. 
%
The condition $D_1=0$ generates a series $\omega_{(1,l)}$ of complex roots indexed $l$, whose real parts determine the resonant frequencies and the imaginary parts determine the quality factor $Q$ of the resonances ($Q\approx 1/|\mbox{Im}\,\omega_{(m,l)}|$). 
By using a Newton Raphson method,  we have obtained   $\omega_{(1,1)}/c=(3.69523-i0.60191)$$\mu$m$^{-1}$ and $(\omega/c)_{(1,2)}=(7.02453-i0.57879)$$\mu$m$^{-1}$, 
whose real part is in very good agreement with the spectral position of the peaks. Consistent with
causality, $\mbox{Im}\,\omega_{(m,l)}$ is strictly negative.  
The emission curve for $\rho'=0.4\mu$m shows the peaks corresponding to resonances whose eigenfrequencies are zeroes of the denominator $D_m$ in Eqs. (\ref{am}) and (\ref{bm}) with $\sigma=0$ and with $1\leq m\leq6$ (see table \ref{tabla1}). The values of the damping rates $|\mbox{Im}\,\omega_{(m,1)}|$ decrease  with $m$, resulting in  high $Q$ resonances as frequency is increased.  This fact is confirmed by  the curve for $\rho'=0.4\mu$m in Figure \ref{figura_sg}a, since the width at half maximum of resonance peaks are decreasing with increasing frequency. 

%
\begin{table}[htbp]
	\centering
\begin{tabular}{c  c  c }
$m$    &    $\mbox{Re}\,\omega_{(m,l)}/c$    &   $|\mbox{Im}\,\omega_{(m,l)} / c|$	\\	
\hline
1 &  1.37324 & 1.75516 \\
\hline	
2 &  3.73355 & 1.83205 \\
\hline
3 &  5.69079 & 0.67741 \\
\hline
4 &  6.86403 & 0.39975 \\
\hline
5 & 8.09922  & 0.24052 \\
\hline
6 & 9.33293  &  0.14277
\end{tabular}
	\caption{Resonance frequencies $\omega_{(m,l)}$ for the first six eigenmodes ($1\leq m \leq 6$ and $l=1$), $R=0.5\mu$m, $\mu_c=0.5$eV, $\gamma=0.1$meV, $\varepsilon_1=3.9$, $\mu_1=1$, $\varepsilon_2=1$, $\mu_2=1$.} 
	\label{tabla1}
\end{table}
%

On the other hand, 
similar values between real parts of the dipolar  $\omega_{(1,2)}/c$ and of the octupolar  $\omega_{(4,1)}/c$ eigenfrequencies, suggest that these two resonances can be excited at the same emission frequency. 
This fact is showed in Figure \ref{figura_sg}a, where two peaks near $7\mu$m$^{-1}$, one on the curve for  $\rho'=0$ and the other one on the curve for $\rho'=0.4\mu$m, are marked with a dot. The insets show the spatial distribution of the magnetic field $\varphi(\rho,\phi)$ calculated at the emission frequency $\omega/c=7\mu$m$^{-1}$ for the case in which the emitter is localized at the wire center (left) and in case where the emitter is localized at $\rho'=0.4\mu$m (right).  This result, is a clear confirmation that different resonances (dipolar and octupolar in this case) can be excited by moving the emitter from the wire center.

%
In Figure \ref{figura_sg}b the normalized spontaneous emission curves are depicted for different orientation angles $\alpha=0,\,30^\circ,\,60^\circ,\,\mbox{and}\,90^\circ$ when the source is placed at $\rho'=0.4\mu$m.  
To illustrate the effects of varying the orientation angle $\alpha$ in the field near to the dielectric wire,  in  Figure \ref{figura_sg}c  we have plotted  the spatial distribution of the magnetic field for the octupolar resonance and for $\alpha=30^\circ, 60^\circ\,\mbox{and}\,90^\circ$. We observe that these shapes are rotated and highly distorted with respect the shape calculated for $\alpha=0$ showed in the inset of Figure \ref{figura_sg}a.


\subsection{Graphene--coated wire}

\begin{figure}
\centering
\resizebox{0.50\textwidth}{!}
{\includegraphics{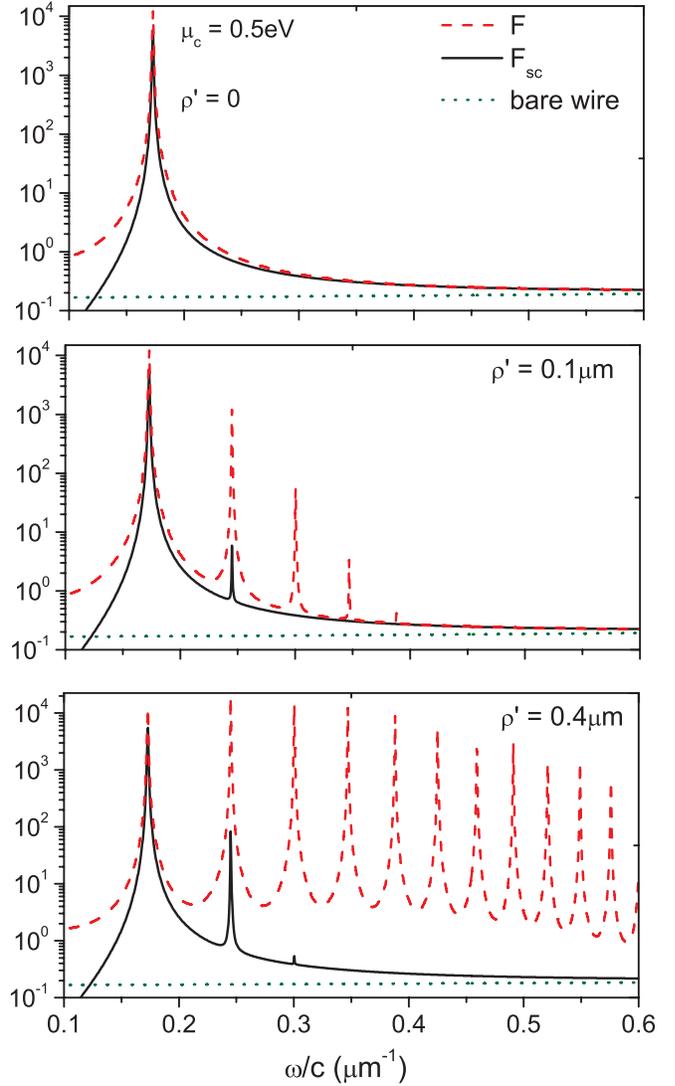}}
\caption{\label{fig:epsart} Efficiency (per unit length) curves,  calculated for $\mu=0.5$eV, $T = 300$ K, $\gamma_c=0.1$meV, $\varepsilon_1=3.9$, $\mu_1=1$ and $\varepsilon_2=1$, $\mu_2=1$. The emission efficiency  curve corresponding to the uncoated cylinder is given as a reference. The emitter is  localized at $\rho'=0$ (a), $\rho'=0.1\mu$m (b), $\rho'=0.4\mu $m (c), with its dipole moment oriented in the $+\hat{x}$ direction ($\alpha=0$). }\label{potencia0p5}
\end{figure}

Having studied bare dielectric wires, we next explore the effects that a graphene coating has on the emission and the radiation spectrum of a dipole emitter localized inside the wire. In particular, the coating is expected to introduce LSPs mechanisms, which were absent in the bare wire, which are able to enhance the emission properties at resonance frequencies. 
%
%
The dashed and solid curves in Figure \ref{potencia0p5} display the frequency dependence of the spontaneous  emission and the
 radiation efficiencies  for the same wire considered in the previous example but  now it is wrapped with a  graphene coating. We used Kubo parameters $\mu_c=0.5$eV, $\gamma=0.1$meV, $T=300$K, emission frequencies in the range between $0.1\mu$m$^{-1}$ ($5$ THz or wavelength $\lambda=60\mu$m) and $6\mu$m$^{-1}$ ($30$ THz or wavelength $\lambda=10\mu$m), and for three different locations of the dipole, $\rho'=0$, 
$0.1\mu$m, and $0.4\mu$m. 
%
Figure \ref{potencia0p5}a shows that both $F$ and $F_{sc}$ efficiencies are enhanced at a frequency near $0.17\mu$m$^{-1}$ ($\lambda=36.35\mu$m) corresponding to the dipolar 
plasmon resonance of the graphene--coated wire. 
The correspondence between the  spectral position of these peaks and  the dipolar plasmon resonance can be clearly seen by
 calculating the complex root $\omega_1/c$ of the common denominator $D_1(\omega)$ in equations (\ref{am}) and (\ref{bm}). We have obtained 
$\omega_1/c=(0.17287-4.57289\,10^{-4})\mu$m$^{-1}$ whose real part is in very good agreement with the spectral position of the peaks.  In contrast to bare wires, where the emission decay rate reaches values close to unity at multipolar resonances,  
we obtain enhancement factors in both the emission and radiation decay rate efficiencies of around
five orders of magnitude when the wire is wrapped with a graphene coating. Furthermore, we find that high--order plasmon resonances are  excited when the emitter is moved from the wire center,  as can be seen in Figure \ref{potencia0p5}b where the curves of $F$ and $F_{sc}$ have been plotted for $\rho'=0.1\mu$m. 
In addition to the maximum enhancement in both the emission and the radiation spectra  at a frequency  near $0.17\mu$m$^{-1}$ ($m=1$ or dipolar resonance), a local maximum enhancement near $0.24\mu$m$^{-1}$  corresponding to the quadrupolar plasmon resonance ($m=2$) appear. Other local maxima, particularly noticeable in the emission  spectra, occur at frequencies near $0.30\mu$m$^{-1}$ ($m=3$ or hexapolar resonance) and $0.35\mu$m$^{-1}$ ($m=4$ or octupolar resonance). 
By  calculating the complex root $\omega_m$ of the common denominator $D_m(\omega)$ in equations (\ref{am}) and (\ref{bm}),
we have obtained  
$\omega_2/c=(0.24509-2.53123\,10^{-4})\mu$m$^{-1}$, $\omega_3/c=(0.30038-2.52376\,10^{-4})\mu$m$^{-1}$ and $\omega_4/c=(0.34697-2.52553\,10^{-4})\mu$m$^{-1}$  
for which their real parts are  in very good  agreement with the spectral position of the peaks. When $\rho'$ is increased, the enhancement factors in the spontaneous emission curve at frequencies corresponding to high--order resonances  becomes more and more pronounced. This fact can be viewed in figure \ref{potencia0p5}c in which the  emission efficiencies for $m=1,\,2,\,3 \, \mbox{and} \,4$ reach similar values ($\approx 10^4$). 

It is worth noting that the width at half maximum of the resonance peaks in Figure \ref{potencia0p5}c, \textit{i.e.}, the quality factor of the resonance $Q$, almost  not shows any significant dependence on the position of the peak. This agree well with the fact that  the damping rate take the value $\mbox{Im}\,\omega_m/c \approx \gamma/2=2.5\,10^{-4}\mu$m$^{-1}$ for large enough $m$ values \cite{CRD}.  
Although   the enhancement factors in the radiation curve are less noticeable than those in the emission curve, these peaks are increased when $\rho'$ is increased, as can be seen by comparing the radiation curves in Figures \ref{potencia0p5}b and \ref{potencia0p5}c.
Moreover, we observe that a peak at a frequency near $0.30\mu$m (in which hexapolar order resonance occur), absent in the curve corresponding to  $\rho'=0.1 \mu$m, becomes just visible.  

\begin{figure}
\centering
\includegraphics[width=\linewidth]{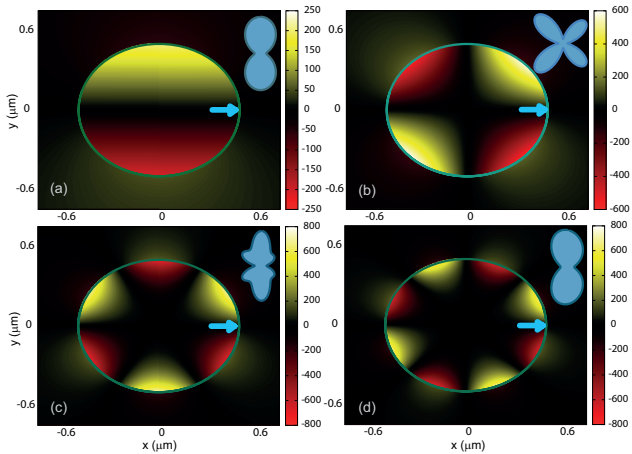}
\caption{\label{fig:epsart} Map of the magnetic field $\varphi(\rho,\phi)$ at a fixed time for plasmon resonances of the wire considered in Figure \ref{potencia0p5}. Red, negative values, yellow, positive  values. The frequency emission is $\omega/c = 0.17287 \mu$m$^{-1}$ (a), $\omega/c = 0.24509 \mu$m$^{-1}$ (b), $\omega/c =0.30038\mu$m$^{-1}$ (c), and $\omega/c=0.34697\mu$m$^{-1}$ (d). The dipole emitter is indicated with an arrow.  
All parameters are the same as in Figure \ref{potencia0p5}c.} \label{campos}
\end{figure}

In Figure \ref{campos} we plot the spatial distribution of the near magnetic field $\varphi(\rho,\phi)$  
for the wire considered in Figure \ref{potencia0p5} and for the first four plasmon resonances.  We have considered the particular case of the emitter localized at $\rho'=0.4\mu$m as in Figure \ref{potencia0p5}c.
%
The emission frequencies are $\omega/c = 0.17287 \mu$m$^{-1}$ (Figure \ref{campos}a), $\omega/c = 0.24509 \mu$m$^{-1}$ (Figure \ref{campos}b), $0.30038\mu$m$^{-1}$ (Figure \ref{campos}c), and $0.34697\mu$m$^{-1}$ (Figure \ref{campos}d), \textit{i.e.}, the values corresponding to the real part of the complex poles $\omega_{(m,1)}/c$ of the multipole coefficients $a_m$ and $b_m$ ($m=1,\,2,\,3,\,\mbox{and}\,4$) and for which the strongest maxima in the emission  efficiency occur. At these frequencies, the near field distributions follow the typical dipolar, quadrupolar, hexapolar and octupolar patterns. Contrary to what happens in bare wires,  the presence of a surface current density $j_\phi= \frac{4\pi \sigma}{c} E_\phi$ induced on the graphene coating and the boundary condition in Eq. (\ref{cc2}) lead to a discontinuity at $\rho=a=0.5 \mu$m as can be  observed in Fig. \ref{campos}. 
For example, the map of the magnetic field showed in Figure \ref{campos}c is characterized by six zones, three of them are red  inside the cylinder and are yellow outside the cylinder, pointing out that the magnetic field  changes from negative values to positive values when passing from inside to outside of the cylinder.  In the other three zones,  yellow  inside the cylinder and  red outside the cylinder, 
the magnetic field  changes from positive to negative values when passing from inside to outside of the cylinder. 

As well as the near field maps at resonance frequencies are closely related with the amplitude and location of the maxima in the emission spectra, the far field maps are related with the peak characteristics in the radiation spectra.   
The inset in Figure \ref{campos}a shows that, for the emission frequency  $\omega/c=0.17287 \mu$m$^{-1}$ ($\omega/c=$Re$ \omega_1/c$), the far field intensities, namely  $d P_{sc}/d \phi$ which is the fraction of the emitted power that is
scattered into the angular region about the scattering direction $\phi$, are dominated by an dipole pattern.   
The integral of $d P_{sc}/d \phi$ over all angles, \textit{i.e.}, the radiative decay rate, is found to be $10^4$ times larger than that of the same dipole emitter in the absence of the graphene--coated wire,  as can be seen in Figure \ref{potencia0p5}c. 
When the emission frequency is $0.24509 \mu$m$^{-1}$ ($\omega/c=$Re$ \omega_2/c$), the radiation efficiency showed in Figure \ref{potencia0p5}c exhibits a strong enhancement (an enhancement factor near three orders
of magnitude greater than that of the same emitter in the absence of the graphene--coated wire) and the far field intensities  are dominated by an electromagnetic quadrupole pattern, as indicated in the inset in figure \ref{campos}b. Similarly, when the emission frequency is $0.30038\mu$m$^{-1}$ ($\omega/c=$Re$ \omega_3/c$), the radiation efficiency plotted in Figure \ref{potencia0p5}c exhibits a very weak peak, and as a result, a small hexapolar pattern overlaps the dipole pattern in the far--field intensities,  
 as indicated in the inset in figure \ref{campos}c. 
On the other hand,  no peak is observed in the radiation efficiency curve when the emission frequency is $0.34697\mu$m$^{-1}$ ($\omega/c=$Re$ \omega_4/c$).  
At this frequency, the far--field intensities are dominated by an electromagnetic  dipole, as indicated in the inset in Figure \ref{campos}d. 

\begin{figure}
\centering
\resizebox{0.450\textwidth}{!}
{\includegraphics{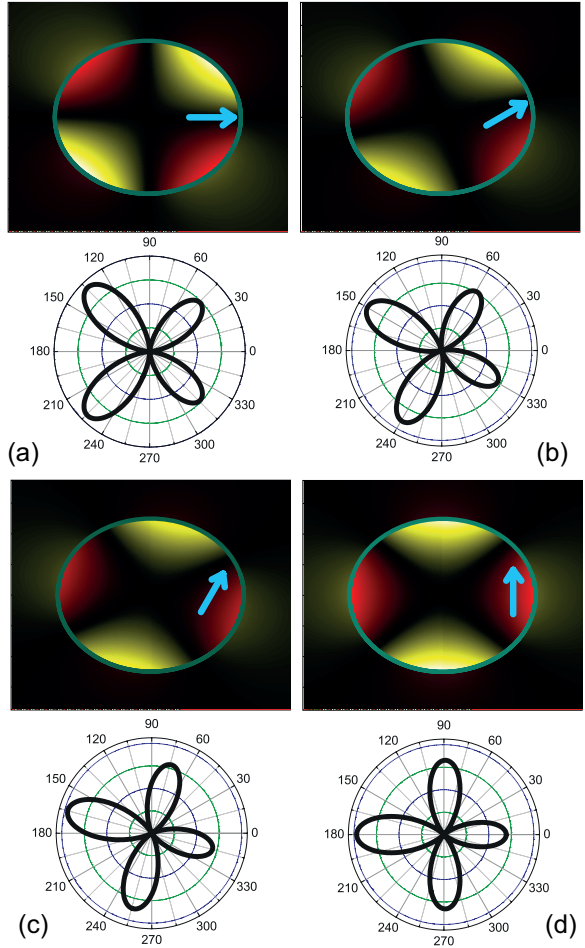}}
\caption{\label{fig:epsart} Map of the near magnetic field $\varphi(\rho,\phi)$ and the far field intensities at a fixed time for the quadrupolar plasmon resonance of the wire and for $\alpha=0$ (a), $\alpha=30^\circ$ (b), $\alpha=60^\circ$, $\alpha=90^\circ$ (d). Red, negative values, yellow, positive  values. The frequency  emission is  $\omega/c = 0.24509 \mu$m$^{-1}$ and the emitter is localized at $\rho'=0.4 \mu$m, $\phi'=0$. The dipole emitter is indicated with an arrow. All other parameters are the same as in Figure \ref{campos}b. } \label{resumen}
\end{figure}

%
Contrary to the case of bare wires, where both the emission and radiation efficiencies are highly modified when the dipole is rotated, in the graphene coated dielectric wire we have observed that neither $F$ nor $F_{sc}$ efficiencies  show  a noticeable dependence on the orientation angle $\alpha$. 
However, the shape of the near field  distribution  and of the far field intensities strongly depend on $\alpha$. This fact can be seen in Figure \ref{resumen}, where we have plotted the near field and the power patterns for  four dipole
orientations: $\alpha=0,\,30^\circ,\,60^\circ\,\mbox{and}\,90^\circ$, at a specifically frequency corresponding to the  quadrupolar plasmon resonance.   
%
When the dipole is rotated,  
both the near field map and the far field intensities are rotated 
with respect the $\alpha=0$ case. Unlike the case of bare  wires, where the near field  distortion due to the rotation of the source is remarkable (see Figure \ref{figura_sg}c), Figure \ref{resumen} shows that when the dielectric wire is  wrapped with a graphene coating only a rotation of the near field is appreciable, being its  distortion  negligible. On the other hand, as $\alpha$ is increased, the shape of the far field intensities are more and more distorted until arriving at the shape shown in Figure \ref{resumen}d corresponding to $\alpha=90^\circ$. Same behavior has been observed [not shown in Figure \ref{resumen}] at   the other  plasmon resonance frequencies.

\begin{figure}
\centering
\resizebox{0.40\textwidth}{!}
{\includegraphics{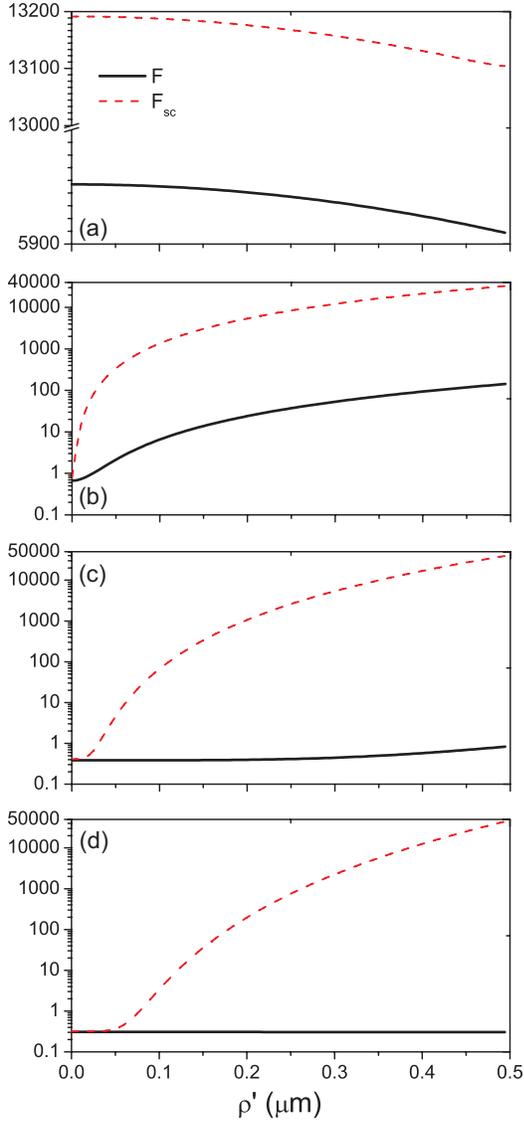}}
\caption{\label{fig:epsart} Emission  and radiation decay efficiencies (per unit length) as a function of the distance from the wire center for the dipole (a), the quadrupole (b), the hexapole (c) and the octupole (d) resonance frequencies. All other the parameters are the same as in Figure \ref{potencia0p5}. The emitter is located on the $\hat{x}$ axis with its dipole moment oriented in the radial direction ($\alpha=0$).} \label{potencia0p5rp}
\end{figure}

To examine the effects of the emitter localization on plasmon resonances, Figure \ref{potencia0p5rp} shows the emission and the radiation decay rates as a function of the distance to the wire center. The emission frequency is chosen such that the emission is resonant at the orders: dipolar (a), quadrupolar  (b), hexapolar  (c) and octupolar (d).  At distance $0\mu$m, the emission  and the radiation decay rates  for the dipole resonance are both $10^4$ times larger than in the absence of the graphene--coated cylinder. When $\rho'$ is increased from this value,  both efficiency curves decrease to reach their lowest value at $\rho'=a=0.5\mu$m as its shown in Figure \ref{potencia0p5rp}a. In contrast, for the quadrupole   resonance,  Figure \ref{potencia0p5rp}b shows that  both efficiency curves take a value close to unity at $\rho'=0\mu$m and that these curves increase with the increasing of $\rho'$. The same behavior is observed in Figure \ref{potencia0p5rp}c for the hexapole resonance, although the maximum value in the radiative decay efficiency  is much lower than the maximum value in the quadrupolar  resonance curve. For the octupole resonance, the value of the radiative decay efficiency does not shows any significant dependence on the position of the emitter, 
as can be seen in  Figure \ref{potencia0p5rp}d.

\section{Conclusions} \label{conclusiones}

In conclusion, we have presented an analytical classical method based on the separation of variables approach to find  the emission and radiation characteristics of an optical emitter inside a circular cross--section wire  coated with a graphene layer.  The method has been used to investigate how the location of the emitter, as well as the orientation of its dipole moment, affect the power emitted by the source. 
We found that the interplay between the optical emitter and the LSPs in the graphene coating strongly influences the spontaneous decay rate as well as  the radiation characteristics. 
We have compared the results with those obtained for the same dielectric wire but without a graphene layer (bare wire), in which the decay rates are enhanced at resonances associated with cavity modes. 

%
%
As might be expected from symmetry arguments, for the optical emitter placed on the wire center, we  found  that both the emission and radiation decay rates are  enhanced just at dipolar resonance frequency.  Instead, an enhancement at frequency of high--order resonances appear when the emitter is moved from the axis of the wire.   
The correspondence between the position of the spectral peaks and the multipolar resonances (cavity or plasmonic) has been clearly shown by calculating the complex poles of the coefficients of the multipole expansion of the electromagnetic field. In case of bare wires, cavity effects are intensified as frequency increases, \textit{i.e.}, where high $Q$ resonances take place. On the contrary, for graphene coated wire,  the resonance damping rate  take the quasistatic value $\mbox{Im}\,\omega_m \approx \gamma/2$ for large enough  $m$ values. As a consequence,  the resonance $Q$ factor  shows any dependence on the position of the spectral peaks. 
Moreover, whereas the emission and radiation decay rates are highly modified with the orientation of the dipole emitter in case of bare wires, we have found that   neither emission nor radiation decay rates show a noticeable dependence on the orientation angle in case of graphene coated wires.

The spatial distribution of the electromagnetic field near the wire for different resonance frequencies and positions of the emitter have been investigated. The multipolar order revealed by the topology of the near field agrees well with the multipolar order revealed by the spectral position of the emission decay rate peak. 
We have shown the modifications in the near field distribution by varying the orientation of the dipole. 



\section*{Acknowledgment}
The author acknowledge the financial support of Consejo Nacional de Investigaciones Cient\'{\i}ficas y T\'ecnicas, (CONICET, PIP 451). 

\section{Appendix I}

In an unbounded medium (constitutve parameters $\varepsilon$, $\mu$) the vector potential $\vec{A(\rho,\phi)}$ satisfy the inhomogeneous Helmholtz equation \cite{novotny}
\begin{eqnarray}\label{helmholtz}
\nabla^2 \vec{A}+k^2 \vec{A}=4\pi i k_0 \vec{p} \frac{1}{\rho} \delta(\rho-\rho') \delta(\phi-\phi')
\end{eqnarray}
where $\rho'$ and $\phi'$ denotes the position of the line dipole source, with $\rho'<a$, $k^2=\varepsilon\,\mu k_0^2$ and $\vec{A}=A_\rho \hat{r}+A_\phi \hat{\phi}$. Since
\begin{eqnarray}\label{delta}
\delta(\phi-\phi')=\frac{1}{2\pi}\sum_{-\infty}^{+\infty} e^{i m (\phi-\phi')}, 
\end{eqnarray}
we can write 
\begin{eqnarray}\label{atau}
A_\tau(\rho,\phi,\rho',\phi') = \sum_{m=-\infty}^{+\infty} a_{\tau.m}(\rho,\rho') e^{i m (\phi-\phi')},
\end{eqnarray}
where the subscript $\tau$ denotes $\rho$ or $\phi$. To determine $a_{\tau,m}(\rho,\rho')$, we insert  expressions (\ref{delta}) and (\ref{atau}) into Eq. (\ref{helmholtz}) and obtain
\begin{eqnarray}\label{helmholts2}
\left[\frac{d}{d \rho} \left(\rho \frac{d}{d \rho}\right) - \frac{m^2}{\rho}+k^2 \rho \right] a_{\tau,m}(\rho,\rho')=\\ \nonumber 
2 i k_0 p_\tau \delta(\rho-\rho'). 
\end{eqnarray}
Taking into account the finiteness at the origin and  the radially outgoing property, we obtain
\begin{eqnarray}\label{helmholts2}
a_{\tau,m}(\rho,\rho')=A J_m(k\rho_<) H_m^{(1)}(k\rho_>),
\end{eqnarray}
where $\rho_<$ ($\rho_>$) is the smaller (larger) of $\rho$ and $\rho'$. The coefficient $A$ is determined by the condition
\begin{eqnarray}\label{helmholts2}
\rho \frac{d a_{\tau,m}}{d \rho}|_{\rho_+}-\rho \frac{d a_{\tau,m}}{d \rho}|_{\rho_-} = 
\frac{2 i k_0 p_\tau}{\rho'}.
\end{eqnarray}
By using $W(J_m(x),H_m^{(1)}(x))= 2 i/\pi x$ where $W$ denotes the Wronskian, we obtain $A=\pi k_0 p_\tau$. Finally,
\begin{eqnarray}
A_\tau(\rho,\phi,\rho',\phi')=\sum_{m=-\infty}^{+\infty} \pi k_0 p_\tau J_m(k\rho_<) H_m^{(1)}(k\rho_>) \nonumber \\ 
\times e^{i m (\phi-\phi')}.
\end{eqnarray} 

\section{Appendix II}
Here, we derive an analytical expression for the first term in Eq. (\ref{P1}), \textit{i.e.}, the contribution of the primary field emitted by the dipole. It is given by 
\begin{eqnarray}\label{P3}
P_i= 
\frac{ \rho_0\,L\,c^2}{8 \pi \omega \varepsilon_1} \int_0^{2 \pi} \mbox{Re} \left\{ -i\, \varphi_i^{*} \frac{\partial}{\partial \rho}\varphi_i \right\} d\phi
\end{eqnarray}
Introducing Eq. (\ref{inc2}) into Eq. (\ref{P3}) we obtain
\begin{eqnarray}
\mbox{Re} \{ \frac{-i\pi k_0^3 k_1 L \rho_0}{8} \sum_{m,n} J_m(x')J_n(x')^* \\ \nonumber
\times \int_0^{2 \pi} [ H_m^{(1)}(x)'' H_n^{(1)}(x)'^* \left|p_\phi \right|^2 \\ \nonumber
-i m \frac{d }{d x}\left(\frac{H_m^{(1)}(x)}{x}\right)H_n^{(1)}(x)'^* p_\rho p_\phi^*\\ \nonumber
+i n H_m^{(1)}(x)'' 
\frac{H_m^{(1)}(x)^*}{x}p_\phi p_\rho^* \\ \nonumber
+m n \frac{d }{d x}\left(\frac{H_m^{(1)}(x)}{x}\right) \frac{H_m^{(1)}(x)^*}{x} \left|p_\rho \right|^2 ] \\ \nonumber
\times e^{i(m-n)(\phi-\phi')} d\phi \},
\end{eqnarray}
where $x=k_1 \rho_0$ and $x'=k_1 \rho'$.
Taking into account  the recurrence relations of the Hankel function \cite{abramowitz}, $p_\rho=p \sin(\alpha-\phi)$ and  $p_\phi=p \cos(\alpha-\phi)$, we can write  Eq. (\ref{P3}) as
%
%
\begin{eqnarray}\label{P4}
P_i=  \frac{c \pi k_0^3 L p^2}{4} 
 \sum_{m=-\infty}^{+\infty} \{ [J_m(x')]^2 \\ \nonumber
+ J_m(x') J_{m+2}(x') \cos(2 [\phi'-\alpha]) \}=\frac{c \pi k_0^3 L p^2 }{4}
\end{eqnarray}
where we have used $W(J_m(x),Y_m(x))=2 / \pi x$, $[J_0(x)]^2+2 \sum_{k=1}^{+\infty} [J_k(x)]^2=1$ and $-J_1(x)^2+2 \sum_{k=0}^{+\infty} J_k(x) J_{k+2}(x)=0$ \cite{abramowitz}. 

It is worth noting that the same result can be obtained by using the Poynting theorem \cite{cuevas1},
\begin{eqnarray}
P_i=\frac{\omega}{2}\mbox{Im}\left\{\vec{p} \cdot \vec{E}_i(\vec{x'}) \right\}
\end{eqnarray}
where the field $\vec{E}_i(\vec{x'})$ is the primary electric dipole field  evaluated at the dipole position $\vec{x'}$.

\end{document}